\documentstyle[12pt]{article}
\begin{document}
\baselineskip 0.65cm
\hyphenation{re-nor-mal-iza-tion-group}
\newcommand{\KEV}{ {\rm keV} }
\newcommand{\MEV}{ {\rm MeV} }
\newcommand{\GEV}{ {\rm GeV} }
\newcommand{\TEV}{ {\rm TeV} }
\newcommand{\gsim}{ \mathop{}_{\textstyle \sim}^{\textstyle >} }
\newcommand{\lsim}{ \mathop{}_{\textstyle \sim}^{\textstyle <} }
\newcommand{\mgut}{M_{\rm GUT}}
\newcommand{\msigma}{M_{\Sigma}}
\newcommand{\mhc}{M_{H_c}}
\newcommand{\mhcbar}{M_{\overline{H}_c}}
\newcommand{\mglu}{m_{\tilde{g}}}
\newcommand{\mwino}{m_{\tilde{w}}}
\newcommand{\mup}{m_{\tilde{u}}}
\newcommand{\md}{m_{\tilde{d}}}
\newcommand{\me}{m_{\tilde{e}}}
\newcommand{\mq}{m_{\tilde{Q}}}
\newcommand{\ml}{m_{\tilde{L}}}
\newcommand{\mh}{m_{\tilde{h}}}
\begin{titlepage}

\begin{flushright}
TU-470
\\
\today
\end{flushright}

\vskip 0.35cm
\begin{center}
{\large \bf Limit on the Color-Triplet Higgs Mass in the Minimum
Supersymmetric SU(5) Model}
\vskip 1.2cm
J.~Hisano\footnote
{Fellow of the Japan Society for the Promotion of Science.}
, T.~Moroi$^1$
, K.~Tobe, and T.~Yanagida

\vskip 0.4cm

{\it Department of Physics, Tohoku University,\\
     Sendai 980-77, Japan}

\vskip 1.5cm

\abstract{
In the minimum supersymmetric SU(5) GUT, we derive the upper limit on
the mass of the color-triplet Higgs multiplets as $\mhc\leq 2.4\times
10^{16}~\GEV$ (90 \% C.L.) taking all possible corrections into account
in a renormalization group analysis. If the above upper limit is
compared with a limit on $\mhc$ from the negative search for the proton
decay; $\mhc \geq 2.0\times 10^{16}~\GEV$ (in which effects of the
larger top-quark mass are included), the minimum supersymmetric SU(5) GUT
is severely constrained.
}

\end{center}
\end{titlepage}

%
%
%
%

The supersymmetric grand unified theory (SUSY-GUT) attracts us as a
candidate of physics beyond the standard model (SM). The
phenomenological successes of the SUSY-GUT are not only the gauge
coupling constant unification~\cite{gcc-unif}, but also the
$m_{b}/m_{\tau}$ ratio~\cite{b-tau}. On the other hand, the proton
decay, which is a direct evidence of GUT, has not yet been observed.  In
the minimum SUSY-SU(5) GUT~\cite{NPB193-150}, the exchanges of
color-triplet Higgs multiplets give rise to the most dominant
contribution to the nucleon decay~\cite{NPB197-533} whose amplitudes
depend on an unknown GUT-scale parameter, that is the mass of the
color-triplet Higgs multiplets $\mhc$~\cite{HMY1,nath}. Therefore, the
negative search for the nucleon decay constrains the model strongly, and
hence it is very important to determine the mass $\mhc$ without any
theoretical prejudices in order to check the consistency of this model.

It has been shown in Ref.\cite{HMY2} that the color-triplet Higgs mass
$\mhc$ can be determined in the minimum SUSY-SU(5) model by using only
the low-energy parameters, {\it {i.e.}}, the gauge coupling constants
and the superparticle mass spectrum at the electroweak scale. However,
in the previous analysis~\cite{HMY2} the effects of the two-loop
corrections below the sfermion mass scale ($\simeq 1~\TEV$) and the
one-loop finite threshold corrections of the SUSY-particles have not
been taken into account. In this letter, we improve the analysis by
including these effects as well as by taking the effect of the top-quark
Yukawa coupling~\cite{Hagelin} into account and by using the most recent
experimental data on the gauge coupling constants~\cite{Langacker}. As a
result, we obtain the upper limit on $\mhc$ as $\mhc \leq 2.4 \times
10^{16}~\GEV$ (90\%~C.L.). We have also checked that the correction from
the $\Sigma({\bf 24})$ loop does not change this result much even if
there exists a large mass splitting among the superheavy particles.

Furthermore, the lower limit on $\mhc$ from the negative search for the
proton decay~\cite{HMY1} is also affected by the low-energy experimental
data.  Recently, CDF collaboration reported the evidence of top-quark
production with its mass $174 \pm 10^{+13}_{-12}~\GEV$~\cite{CDF}.
Thus, we need to reanalyze the constraints on $\mhc$ since the top-quark
mass has been assumed at $90~\GEV$ in the previous analysis~\cite{HMY1}.
When this lower limit on $\mhc$ from the negative search for 
the proton decay is compared with the above upper limit on $\mhc$
derived from the renormalization group (RG) analysis, we find that the
minimum SUSY-SU(5) GUT is severely constrained and we conclude that the
minimum SUSY-SU(5) model is consistent only in a very narrow parameter
region. 

Let us start with studying the minimum SUSY-SU(5) model. This model
contains the following chiral supermultiplets;
\begin{eqnarray}
{\rm matter}:~\psi_{i}({\bf 10}),~~~\phi_{i}({\bf \overline{5}}),~~~~~
{\rm Higgs}:~H({\bf 5}),~~~\overline{H}({\bf \overline{5}}),
~~~\Sigma({\bf 24}),
\label{contents}
\end{eqnarray}
where $i$(=1,2,3) represents the family index. The superpotential in
this model is 
\begin{eqnarray}
W &=& \frac{f}{3}Tr\Sigma^3+\frac{1}{2} fV Tr\Sigma^2
	+\lambda \overline{H}_{A}(\Sigma_B^A+3V \delta_B^A)H^B
\nonumber \\ &&
	+\frac{h^{ij}}{4}\epsilon_{ABCDE}\psi_{i}^{AB}\psi_{j}^{CD}H^E
	+\sqrt{2}f^{ij}\psi_{i}^{AB}\phi_{jA}\overline{H}_{B},
\label{superpotential}
\end{eqnarray}
where indices $A,B,C,...$ are the SU(5) indices which run from 1 to 5
and $\epsilon_{ABCDE}$ the fifth-antisymmetric tensor.

The {\bf 24}-dimension Higgs $\Sigma({\bf 24})$ has the following
vacuum-expectation value that causes the breaking SU(5) $\rightarrow
{\rm SU(3)_{C}} \times {\rm SU(2)_{L}} \times {\rm U(1)_{Y}}$,
\begin{eqnarray}
\langle \Sigma \rangle = V
\left(
\begin{array}{ccccc}
        2 & & & & \\
	 & 2 & & & \\
	 & & 2 & & \\
	 & & & -3 & \\
	 & & & & -3 \\
\end{array}
\right).
\label{VEV}
\end{eqnarray}
This vacuum-expectation value gives the following masses to {\it X} and
{\it Y} gauge bosons corresponding to the broken SU(5) generators;
\begin{eqnarray}
M_{V}=M_{X}=M_{Y}=5\sqrt{2}g_{5}V,
\end{eqnarray}
where $g_5$ is the unified SU(5) gauge coupling constant. The invariant
mass parameter of $H({\bf 5})$ and $\overline{H}({\bf \overline{5}})$ is
fine-tuned to realize massless ${\rm SU(2)_{L}}$-doublet Higgs
multiplets $H_{f}$ and $\overline{H}_{f}$ while their color-triplet
partners $H_{c}$ and $\overline{H}_{c}$ are kept superheavy as 
\begin{eqnarray}
\mhc=\mhcbar=5 \lambda V .
\end{eqnarray}
After the SU(5) symmetry breaking, $\Sigma({\bf 24})$ is decomposed into
various mass eigenstates. The masses $M_{\bf 8}$ and $M_{\bf 3}$ for the
components ({\bf 8}, {\bf 1}) and ({\bf 1}, {\bf 3}) under ${\rm
SU(3)_C}\times {\rm SU(2)_L}$ are given by
\begin{eqnarray}
\msigma \equiv M_{\bf 8}=M_{\bf 3}=\frac{5}{2}fV ,
\end{eqnarray}
while the components ({\bf 3}, {\bf 2}) and (${\bf \overline{3}}$, {\bf
2}) form superheavy vector multiplets of mass $M_{V}$ together with the
gauge multiplets {\it X} and {\it Y}, and the mass of singlet component
({\bf 1}, {\bf 1}) is $\frac{1}{2}fV = \frac{1}{5}\msigma$~\cite{HMY1}.

Before performing numerical analyses, let us briefly review the
procedure we use. The masses $M_V$, $\msigma$, and $\mhc$ for the
superheavy particles are constrained from the low-energy parameters,
{\it {i.e.}} the gauge coupling constants and the superparticle mass
spectrum, as shown in Ref.\cite{HMY2}.  Requiring the unification
condition, we can relate three gauge coupling constants at $\mu=m_{Z}$
(with $\mu$ being a renormalization point) to the SU(5) gauge coupling
constant $\alpha_{5}=g_{5}^{2}/4 \pi$ at $\mu=\Lambda \gg
M_{V},~\msigma,$ and $\mhc$. At the one loop level, the relations are 
given by\footnote{The ${\rm U(1)_{Y}}$ gauge coupling constant is
normalized such as $\alpha_{Y}= \frac{3}{5}\alpha_{1}$.}
\begin{eqnarray}
\alpha^{-1}_{3}(m_{Z}) &=& 
	\alpha^{-1}_{5}(\Lambda)
	+\frac{1}{2\pi} 
\left\{
	\left( -2-\frac{2}{3}N_{g} \right) \ln{\frac{m_{\rm SUSY}}{m_{Z}}}
\right.
\nonumber \\
&&
\left.
	+ \left( -9+2N_{g} \right) \ln{\frac{\Lambda}{m_{Z}}}
	-4\ln{\frac{\Lambda}{M_{V}}}
	+3\ln{\frac{\Lambda}{\msigma}}
	+\ln{\frac{\Lambda}{\mhc}}
\right\},
\label{al3}
\\
\alpha^{-1}_{2}(m_{Z}) &=& 
	\alpha^{-1}_{5}(\Lambda)
	+\frac{1}{2\pi} 
\left\{
	\left( -\frac{4}{3}-\frac{2}{3}N_{g}-\frac{5}{6} \right)
	\ln{\frac{m_{\rm SUSY}}{m_{Z}}}
\right.
\nonumber \\
&&
\left.
	+ \left( -6+2N_{g}+1 \right) \ln{\frac{\Lambda}{m_{Z}}}	
	-6\ln{\frac{\Lambda}{M_{V}}}
	+2\ln{\frac{\Lambda}{\msigma}}
\right\},
\label{al2}
\\
\alpha^{-1}_{1}(m_{Z}) &=& 
	\alpha^{-1}_{5}(\Lambda)
	+\frac{1}{2\pi} 
\left\{
	\left( -\frac{2}{3}N_{g}-\frac{1}{2} \right)
	\ln{\frac{m_{\rm SUSY}}{m_{Z}}}
\right.
\nonumber \\
&&
\left.
	+ \left( 2N_{g}+\frac{3}{5} \right) \ln{\frac{\Lambda}{m_{Z}}}
	-10\ln{\frac{\Lambda}{M_{V}}}
	+\frac{2}{5}\ln{\frac{\Lambda}{\mhc}}
\right\},
\label{al1}
\end{eqnarray}
where $N_g$ represents the number of the families. In Eqs.(\ref{al3})
-- (\ref{al1}) we have assumed that all superparticles in the
SUSY-standard model have a common SUSY-breaking mass, $m_{\rm SUSY}$,
for simplicity. By taking the suitable linear combinations of
Eqs.(\ref{al3}) -- (\ref{al1}), we obtain simple relations,
\begin{eqnarray}
\left(
	3\alpha^{-1}_{2}-2\alpha^{-1}_{3}-\alpha^{-1}_{1}
\right)(m_{Z}) &=&
	\frac{1}{2\pi}
\left\{
	\frac{12}{5} \ln \frac{\mhc}{m_{Z}}
	-2 \ln \frac{m_{\rm SUSY}}{m_{Z}}
\right\},
\label{colormass} \\
\left(
	5\alpha^{-1}_{1}-3\alpha^{-1}_{2}-2\alpha^{-1}_{3}
\right)(m_{Z}) &=&
	\frac{1}{2\pi}
\left\{
	12 \ln \frac{M_{V}^{2} \msigma}{m_{Z}^{3}}
	+8\ln \frac{m_{\rm SUSY}}{m_{Z}}
\right\}.
\label{gutmass}
\end{eqnarray}
These equations imply that we can give the constraint to the superheavy
masses from the precision data of $\alpha_{1}$, $\alpha_{2}$ and
$\alpha_{3}$ at the electroweak scale. 

In a qualitative analysis, the limit on $\mhc$ depends on the mass
spectrum of the superparticles. The effect of the mass splittings among
superpartners can be taken into account by replacing $\ln \left
( {m_{\rm SUSY}}/{m_{Z}} \right)$ in Eq.({\ref{colormass}}) as
\begin{eqnarray}
-2\ln \frac{m_{\rm SUSY}}{m_{Z}}~~ \rightarrow ~~
	4\ln \frac{\mglu}{\mwino}
	+\frac{N_g}{5} \ln
	\frac{{\mup}^{3}{\md}^{2}{\me}}{{\mq}^{4}{\ml}^{2}}
	-\frac{8}{5}\ln \frac{\mh}{m_{Z}}
	-\frac{2}{5}\ln \frac{m_{H}}{m_{Z}},
\label{colorthreshold}
\end{eqnarray}
and that in Eq.(\ref{gutmass}) as 
\begin{eqnarray}
8\ln \frac{m_{\rm SUSY}}{m_{Z}}~~ \rightarrow ~~
	4 \ln \frac{\mglu}{m_{Z}}
	+4 \ln \frac{\mwino}{m_{Z}}
	+ N_{g} \ln \frac{{\mq}^{2}}{\me \mup}.
\label{gutyhreshold}
\end{eqnarray}
Here, two doublet Higgs bosons are assumed to have masses at $m_{H}$ and
$m_{Z}$, respectively. The symbols $m_{\tilde q}~ ({\tilde q}={\tilde
Q},~{\tilde u},~{\tilde d})$, $m_{\tilde l}~({\tilde l}={\tilde
L},~{\tilde e})$, $\mwino$, $\mglu$, and $\mh$ represent the masses of
squarks, sleptons, wino, gluino and doublet Higgsinos. From
Eq.(\ref{colormass}) and Eq.(\ref{colorthreshold}), one can see that the
$\mhc$ increases as $\mh$ and $m_{H}$ become larger. In order to derive
a conservative upper limit on $\mhc$, we should use the maximal values
of $\mh$ and $m_{H}$ allowed from the naturalness point of view.  In our
analysis, we take $\mh = m_{H} = 1 \TEV$. The mass ratio $\mglu/\mwino$
is given by $\alpha_{3}/\alpha_{2}$ because of the unification of the
gaugino masses at the GUT scale, and hence the first term in the
right-handed side of Eq.(\ref{colorthreshold}) gives a constant
independent of the gaugino masses. Assuming that all the sfermion masses
are universal at the GUT scale, the second term takes the minimal value
if the gaugino mass is much smaller than the sfermion masses (see
Ref.\cite{HMY1,nojiri}). Thus, the case where the sfermion masses are
heavier than the gaugino masses gives the conservative upper limit on
$\mhc$ at the one-loop level. Though we have assumed that the sfermion
masses for $\phi(\overline{\bf 5})$ and $\psi({\bf 10})$ are the same at
the GUT scale, they may be different each other. In the situation that
the sfermion masses for $\psi({\bf 10})$ are comparable to gaugino
masses at the electroweak scale while the ones for $\phi(\overline{\bf
5})$ are heavier than them, the upper limit on $\mhc$ is raised by
factor 1.6.  However, this choice of the mass spectrum makes the proton
lifetime much shorter, and hence we do not consider this situation. From
these arguments, we take the wino mass smaller than 200~GeV,\footnote
{In the case of $m_{\tilde{h}}\gg m_W, m_{\tilde{w}}$, mixings among
gauginos and doublet Higgsinos can be neglected.}
and all sfermion masses at $1~\TEV$. It
should be noted here that the negative search for the proton decay
prefers this situation.

As the input parameters, we use the most recent values of the
${\overline {\rm MS}}$ (a modified minimal subtraction) gauge coupling
constants at the {\it Z}-pole given in Ref.{\cite{Langacker}},
\begin{eqnarray}
\nonumber
\left. \alpha^{-1}(m_{Z}) \right|_{\rm SM} &=&127.9 \pm 0.1,
\\
\nonumber
\left. \sin^{2} \theta_{W}(m_{Z}) \right|_{\rm SM}
&=& 0.2317 \pm 0.0003 \pm 0.0002,
\\
\left. \alpha_{3}(m_{Z}) \right|_{\rm SM} &=& 0.116 \pm 0.005,
\label{experimental_data}
\end{eqnarray}
where the index SM means that these values are defined in the framework
of the SM. Here, the second error in the Weinberg angle arises from the
ambiguity of the Higgs mass ($m_h$) of the SM, and it has been taken
from $60~\GEV~(-)$ to $1~\TEV~(+)$. In the minimum supersymmetric
standard model (MSSM) the mass of the SM-like Higgs is expected to be
small ($\leq 150~\GEV$~\cite{OYY}). Thus, we use the Weinberg angle
corresponding to the case of $m_{h}=60~\GEV$ in the present
analysis.\footnote
{When one uses the Weinberg angle with $m_{h}=300~\GEV$, the upper limit
on $\mhc$ is raised up only by factor 1.3.}
Since we adopt the $\overline{\rm DR}$-scheme (a dimensional reduction
with the modified minimal subtraction) in our RG analysis, we have to
convert the $\overline{\rm MS}$ coupling constants at the {\it Z}-pole
into the $\overline{\rm DR}$ ones;
\begin{eqnarray}
\frac{1}{\alpha_{i}^{\overline{\rm DR}}(m_{Z})}=
	\frac{1}{\alpha_{i}^{\overline{\rm MS}}(m_{Z})}
 	-\frac{C_{i}}{12 \pi},
\end{eqnarray}
where $C_{1}=0$, $C_{2}=2$ and $C_{3}=3$~\cite{NPB211-216}.

In our numerical calculations, we derive the constraints on $\mhc$ by
using the two-loop RG equations. The decoupling of heavy particles
is taken into account at each mass threshold, namely, for $\mu \geq
m_{\rm SUSY}$ we use the RG equations of the MSSM, for $\mglu
\leq \mu \leq m_{\rm SUSY}$ those of the SM with the wino and the
gluino, and so on. Furthermore, we also include the one-loop finite
threshold corrections to the gauge coupling constants due to the gaugino
loops. Since we assume that the superparticles except for the gauginos
are sufficiently heavy, we only have to consider the finite threshold
corrections from the gaugino loops to the self energies of the gauge 
bosons.\footnote
{In the case that the sfermions, the extra Higgs doublet, and the
Higgsinos are all heavy, the box and vertex corrections are negligible.}
In the case $m_{\tilde{g}_{i}}>m_Z$ ($m_{\tilde{g}_{2}}=\mwino$,
$m_{\tilde{g}_{3}}=\mglu$), we use the following matching
condition~\cite{Yamada} between the gauge coupling constants
$\alpha_{i}$ in the effective theory defined at $\mu >
m_{\tilde{g}_{i}}$ and those at $\mu < m_{\tilde{g}_{i}}$,
\begin{eqnarray}
\alpha_{i}^{-1}(m_{\tilde{g}_{i}} -0) 
&=& \alpha_{i}^{-1}(m_{\tilde{g}_{i}} +0)
\nonumber \\ &&
- \frac{C_{i}}{\pi} \int_{0}^{1} dx~x(1-x)
\ln \left[
1-\frac{m_{Z}^{2}}{m_{\tilde{g}_{i}}^{2}} x(1-x)
\right],
\label{delta1}
\end{eqnarray}
On the contrary, if the wino mass is smaller than the $Z$-boson
mass,\footnote
{In our analysis, we always use the gluino mass larger than the
$Z$-boson mass since the lower limit $\mglu\geq
135~\GEV$ is obtained by the CDF experiments~\cite{cdf_on_mglu}.}
we use the following matching condition~{\cite{Yamada}} for the
SU(2)$_{\rm L}$ gauge coupling constant;
\begin{eqnarray}
\left. \alpha_{2}^{-1}(m_Z) \right|_{\rm SM}
&=& \left. \alpha_{2}^{-1}(m_Z) \right|_{{\rm SM}+\tilde{w}}
\nonumber \\ &&
- \frac{C_{2}}{\pi} \left\{
\frac{1}{3}\ln \frac{\mwino}{m_Z} + 
\int_{0}^{1} dx~x(1-x) \ln\left[
1-\frac{m_{Z}^{2}}{m_{\tilde{w}}^{2}} x(1-x) \right]
\right\},
\label{delta2}
\end{eqnarray}
while those for SU(3)$_{\rm C}$ and U(1)$_{Y}$ are 
$\alpha_i^{-1}(m_Z)|_{\rm SM} = \alpha_i^{-1}(m_Z)|_{{\rm
SM}+\tilde{w}}$ ($i=1,3$). Here, $ \alpha_i^{-1}(m_Z)|_{{\rm
SM}+\tilde{w}}$ are the gauge coupling constants at $\mu=m_{Z}$ defined
in the SM with the wino. Since we now consider the situation where the
gaugino masses are close to $m_{Z}$, these finite threshold corrections
are not negligible.  Notice that the effects of the one-loop finite
threshold corrections and the two-loop corrections below the sfermion
mass scale have not been included in the previous
analysis~{\cite{HMY2}}. Numerically, these corrections decrease the
previous limit on $\mhc$ by factor $\sim 0.5$.

Next, we comment on the effect of the top-quark Yukawa coupling in the
RG equations at the two-loop level. Recently, the CDF collaboration has
announced the evidence for top-quark production and reported $m_t=174
\pm 10_{-12}^{+13}~\GEV$~{\cite{CDF}}.  Thus, since the top-quark Yukawa
coupling constant $y_{t}$ is large, its effect should be
included~{\cite{Hagelin}}. In our numerical analysis, we perform the
full integration of the RG equations at the two-loop level and study the
effect of the top-quark Yukawa coupling.  Here, we take $m_{t}=174~\GEV$
(with $m_{t}$ being the physical mass of the top quark), and vary
$\tan\beta_{H}$ from 1.3 to 5.\footnote
{If $\tan\beta_{H}$ is smaller than $\sim 1.3$, $y_{t}$ blows up 
below the GUT scale, while a large value of $\tan\beta_{H}$ conflicts
with the negative search for the proton decay. Therefore, we have taken
$\tan \beta_{H}=1.3-5$. In this case, the effects of the
bottom-quark and tau-lepton Yukawa coupling constants are not important,
and hence we neglect them below.}
For the case $\tan\beta_{H}=1.3~(2, 5)$, the effect of top-quark Yukawa
coupling raises the upper limit on $\mhc$ by factor 2 (1.3, 1.2).

Combining all the above corrections,\footnote
{Using the $1 \sigma$ errors of the gauge coupling constants in
Eq.(\ref{experimental_data}), the constraint on
$\mhc$ is given by
\begin{eqnarray}
\nonumber
\mhc \leq~ 4.7 \times 10^{15}~\GEV,
\end{eqnarray}
which should be compared with the previous result in Ref.{\cite{HMY2}}
\begin{eqnarray}
\nonumber
\mhc \leq ~ 2.3 \times 10^{17}~\GEV ,
\end{eqnarray}
where the experimental data $\alpha^{-1}(m_{Z})=127.9 \pm 0.2$,
$\sin^{2} \theta_{W}(m_{Z})=0.2326 \pm 0.0008$ and
$\alpha_{3}(m_{Z})=0.118 \pm 0.007$ were used. The main reasons
that the upper limit on $\mhc$ lowers come from the falloff of the
central values of $\sin^{2} \theta_{W}$ and $\alpha_{3}$, and the
smaller error bar. Furthermore, the effects of the two-loop
corrections below the sfermion mass scale and the one-loop finite 
corrections give a smaller value in the final result, which have
not been taken into account in the previous analysis~\cite{HMY2}.}
we calculate the upper limit on $\mhc$. In Fig.~1, the constraints on
$\mhc$ are shown in the $\mwino$ -- $\mhc$ plane. As one can see, the
upper limit decreases steeply when $\mwino$ is smaller than $\sim m_Z$
since the finite correction to the ${\rm SU(2)_{L}}$ gauge coupling
constant given in Eq.(\ref{delta2}) is significant in such a region. For
the case of a heavy wino ($\mwino\gsim m_Z$), the finite correction
given in Eq.(\ref{delta1}) from the wino (and the gluino) loop is
negligible while the RG effects at the two-loop level lower $\mhc$
slightly. As a result, the upper limit obtained at $\mwino =
100~\GEV$ is given by
\begin{eqnarray}
\mhc \leq 2.4\times 10^{16}~\GEV ~~~({\rm 90\%~C.L.}),
\label{upperbound}
\end{eqnarray}
where we have taken $\tan\beta_H = 1.8$.\footnote
{As we will see later, the proton-decay rate have its minimum value at
$\tan\beta_H\simeq 1.8$.}
In Fig.~2, we show the $\tan\beta_{H}$ dependence of the upper limit on
$\mhc$ in the RG analysis, assuming $\mwino=100~\GEV$. It comes from the
effect of the large top-quark Yukawa coupling constant that the upper
limit increases in the small $\tan\beta_{H}$ region. On the other hand,
the upper limit on $\mhc$ is almost independent of $\tan\beta_H$ in the
large $\tan \beta_{H}$ region since the effect of the top-quark Yukawa
coupling is small there. 

The upper limit on $\mhc$ in Eq.(\ref{upperbound}) should be compared
with the lower limit on $\mhc$ derived from the negative search for the
proton decay~\cite{particle-data}.  The amplitude of the nucleon decay
is proportional to the charm- and strange-quark Yukawa coupling
constants at the GUT scale,\footnote
{In order to evaluate the strange-quark Yukawa coupling constant at the
GUT scale, we have used the strange-quark mass at 1GeV,
$m_s=175\MEV$~\cite{strange-mass}. However, if one uses the muon mass
$m_\mu=106\MEV$ instead, one gets almost three times larger amplitude
for the proton decay. From this point also the present limit on $\mhc$
is regarded as a conservative one.}
and it takes its minimal value when $\tan \beta_{H}=1$
(see Ref.{\cite{HMY1}}). However, since the top quark
is very heavy ($m_{t} \simeq 174~\GEV$), the top-quark Yukawa coupling
constant blows up below the GUT scale if $\tan \beta_{H}\lsim 1.3$, and
hence we can not take $\tan\beta_{H}=1$. Moreover, in the RG analysis,
the charm-quark Yukawa coupling constant at the GUT scale becomes larger
due to the renormalization effects by the large top-quark Yukawa
coupling, especially for the case of small $\tan \beta_{H}$.  Therefore,
we reanalyze the nucleon decay amplitudes. (In Ref.\cite{HMY1} the
top-quark mass has been assumed at $90~\GEV$.) In Fig.~2 we also show the
lower limit on $\mhc$ derived from the negative search for the proton
decay in the case of $\mwino=100~\GEV$.  In order to derive a
conservative constraint on $\mhc$, we have taken all sfermion masses at
$1\TEV$, and assumed that the diagrams of scalar-charm and scalar-top
exchanges cancel out in the amplitude of the decay mode $p,~n\rightarrow
K~{\overline{\nu}}_{\mu}$ and that the dominant decay mode is
$p,~n\rightarrow\pi~{\overline{\nu}}_{\mu}$~\cite{HMY1}. The
hadron matrix element\footnote
{The hadron matrix element $\beta$ is defined as
\begin{eqnarray*}
\beta u_L ({\bf k}) \equiv \epsilon_{abc}
\langle 0 | (d_L^a u_L^b) u_L^c | p_{\bf k} \rangle,
\end{eqnarray*}
where $u_L ({\bf k})$ is the wave function of the proton with a momentum
{\bf k}, $u_L$ and $d_L$ are the field operators for the up and
down quarks, and $| p_{\bf k} \rangle$ represents the one-particle state
of the proton with a momentum {\bf k}. Here, $a$, $b$ and $c$ are the
color indices which run 1 -- 3.}
$\beta$ is obtained from the lattice calculation~\cite{lattice}
\begin{eqnarray}
\beta = (5.6 \pm 0.8 ) \times 10^{-3}~\GEV^3.
\label{beta}
\end{eqnarray}
To give a conservative constraint, we have used $\beta = 0.004~\GEV^3$
taking the $2 \sigma$ errors in Eq.(\ref{beta}). Therefore, the
obtained lower limit on $\mhc$ should be regarded as a very conservative
one. From Fig.~2, we read off the lower limit on $\mhc$ as
\begin{eqnarray}
\mhc \geq 2.0\times 10^{16}~\GEV,
\label{lowerbound}
\end{eqnarray}
which corresponds to $\tan\beta_H \simeq 1.8$.
In Fig.~1, we also show the lower limit on $\mhc$ in the $\mwino$ --
$\mhc$ plane. From Figs.~1 and 2, we can see that the minimum SUSY-SU(5)
model is severely constrained, and there remains only a very narrow
parameter region.\footnote
{Note that the proton-decay amplitudes are very sensitive to the details
of the GUT model. Though the minimum SUSY-SU(5) model is very strongly
constrained by the experiments, there is a model in which the
proton-decay rate is suppressed naturally~\cite{TU-461}.}

So far we have assumed that $\msigma \sim M_{V} \sim 10^{16}~\GEV$.
However, the two-loop correction from the physical $\Sigma$-Higgs loop
may loosen the limit on $\mhc$ given in Eq.(\ref{upperbound}) if the 
$\msigma$ is much smaller than $\mhc~(\simeq~10^{16}~\GEV)$.
From the RG analysis, $\msigma$ and $M_{V}$
are constrained only in the combination as $(M_{V}^{2}\msigma)^{1/3}
\simeq 10^{16}~\GEV$, as it has been shown in Ref.{\cite{HMY2}}. Thus,
$\msigma$ and $M_{V}$ may split largely under the above constraint. For
the case $\msigma\gg M_{V}$ ($\msigma
\sim fV $ and $M_{V} \sim g_{5}V$), the Yukawa coupling constant $f$ is
so large that it blows up below the gravitational scale.  We consider,
therefore, the case $\msigma \ll M_{V}$.\footnote
{In this extreme case we can solve the serious Polonyi
problem~\cite{Polonyi} in supergravity as stressed in
Ref.~\cite{TU455}.}
In this case, in the range, $\msigma \leq \mu \leq min(\mhc,~M_{V})$,
the theory is effectively described by the MSSM with the $\Sigma$-Higgs
of a mass $\msigma$. (Here, we use the notation $\Sigma$ as the
components ({\bf 8}, {\bf 1}) and ({\bf 1}, {\bf 3}) in the $\Sigma({\bf
24})$.) Then, the $\Sigma$-loop may raise the upper limit on $\mhc$ at
the two loop level. In the MSSM with the $\Sigma$-Higgs, the two-loop RG
equations for the gauge coupling constants are given by\footnote
{Strictly speaking, there are more contributions from Yukawa couplings
$f Tr{\Sigma^{3}}$ and $\lambda \overline{H} \Sigma H$ in the two-loop
RG equations. But, in the case $\msigma \ll M_{V}$ the Yukawa coupling
constant $f$ is small. We have checked that the Yukawa coupling constant
$\lambda$ is also negligible as far as $\lambda$ stays in the
perturbative regime below the gravitational scale $\simeq
10^{18}~\GEV$.}
\begin{eqnarray}
\mu \frac{d \alpha^{-1}_{i}}{d \mu}=
-\frac{1}{2 \pi} \left[ b_{i}+ \frac{1}{4 \pi}
\left( \sum_{j=1}^{3} b_{ij} \alpha_{j}
	-a_{i t} \alpha_{t} \right)
\right]~~~(i=1,2,3),
\label{RGE-2loop}
\end{eqnarray}
with
\begin{eqnarray}
b_{i}&=&b_{i}^{\rm {MSSM}} +
\left(
\begin{array}{ccc}
0, & 2, & 3
\end{array}
\right)^{\Sigma-{\rm Higgs}},
\\
b_{ij}&=&b_{ij}^{\rm {MSSM}} + 
\left(
\begin{array}{ccc}
0 &    &   \\
  & 24 &   \\
  &    & 54 
\end{array}
\right)^{\Sigma-{\rm Higgs}},
\\
a_{i t}&=& a_{i t}^{\rm {MSSM}},
\end{eqnarray}
where $\alpha_{t}=y_{t}^2/4 \pi$, 
$(....)^{\rm {MSSM}}$ is the contribution of MSSM and
$(....)^{\Sigma-{\rm Higgs}}$ that of $\Sigma$-Higgs.
For the extreme case, $\msigma = 10^{13}~\GEV$ and $M_{V} \simeq
10^{18}~\GEV$, we obtain the limit on $\mhc$ as
\begin{eqnarray}
\mhc \leq 3.7\times 10^{16}~\GEV~~~(90~\%~{\rm C.L.}).
\label{colormassbound2}
\end{eqnarray}
In deriving Eq.({\ref {colormassbound2}}), we have taken $\mwino=100~\GEV$
and $\tan \beta_{H}=1.8$.
These effects raise the upper limit on $\mhc$ only by factor $\sim 1.5$
compared with the result given in Eq.(\ref{upperbound}),\footnote
{We comment on the case when the $\lambda \overline{H} \Sigma H$
coupling constant is large ($\lambda\geq 10$). In this case, we have
found that the constraint in Eq.(\ref{colormassbound2}) is much weakened
and even the larger mass $\mhc \sim 10^{18}~\GEV$ becomes consistent
with the low-energy data on the gauge coupling constants. However, we
need to perform non-perturbative analysis on RG equations in order to
confirm this scenario.}
but the situation does not change much.

In summary, we conclude that the minimum SUSY-SU(5) model is severely
constrained and that there survives only a very narrow parameter
region. The super-KAMIOKANDE experiment will, therefore, give us a
conclusive test on the minimum SUSY-GUT.
\\~\\
\underline{acknowledgement}\\
The authors would like to thank H.~Murayama for useful discussions.
\newpage

%
%
\newcommand{\Journal}[4]{{\sl #1} {\bf #2} {(#3)} {#4}}
\newcommand{\PL}{\sl Phys. Lett.}
\newcommand{\PR}{\sl Phys. Rev.}
\newcommand{\PRL}{\sl Phys. Rev. Lett.}
\newcommand{\NP}{\sl Nucl. Phys.}
\newcommand{\ZP}{\sl Z. Phys.}
\newcommand{\PTP}{\sl Prog. Theor. Phys.}
\newcommand{\NC}{\sl Nuovo Cimento}

\newpage

\section*{Figure Captions}
{\bf Fig.~1}\\ The upper limit on $\mhc$ from the RG analysis and the
lower limit from the negative search for the proton decay as a function
of $\mwino$. The solid line is the upper limit and the dashed line the
lower limit. The shaded region is excluded. Here, we take $m_{\tilde 
q}=m_{\tilde l}=m_{\tilde h}=m_{H}=1~\TEV$, $\beta=0.004~\GEV^{3}$ and
$m_{t}=174~\GEV$. We choose $\tan \beta_{H}=1.8$ in order to
minimize the decay rate of proton.
\\
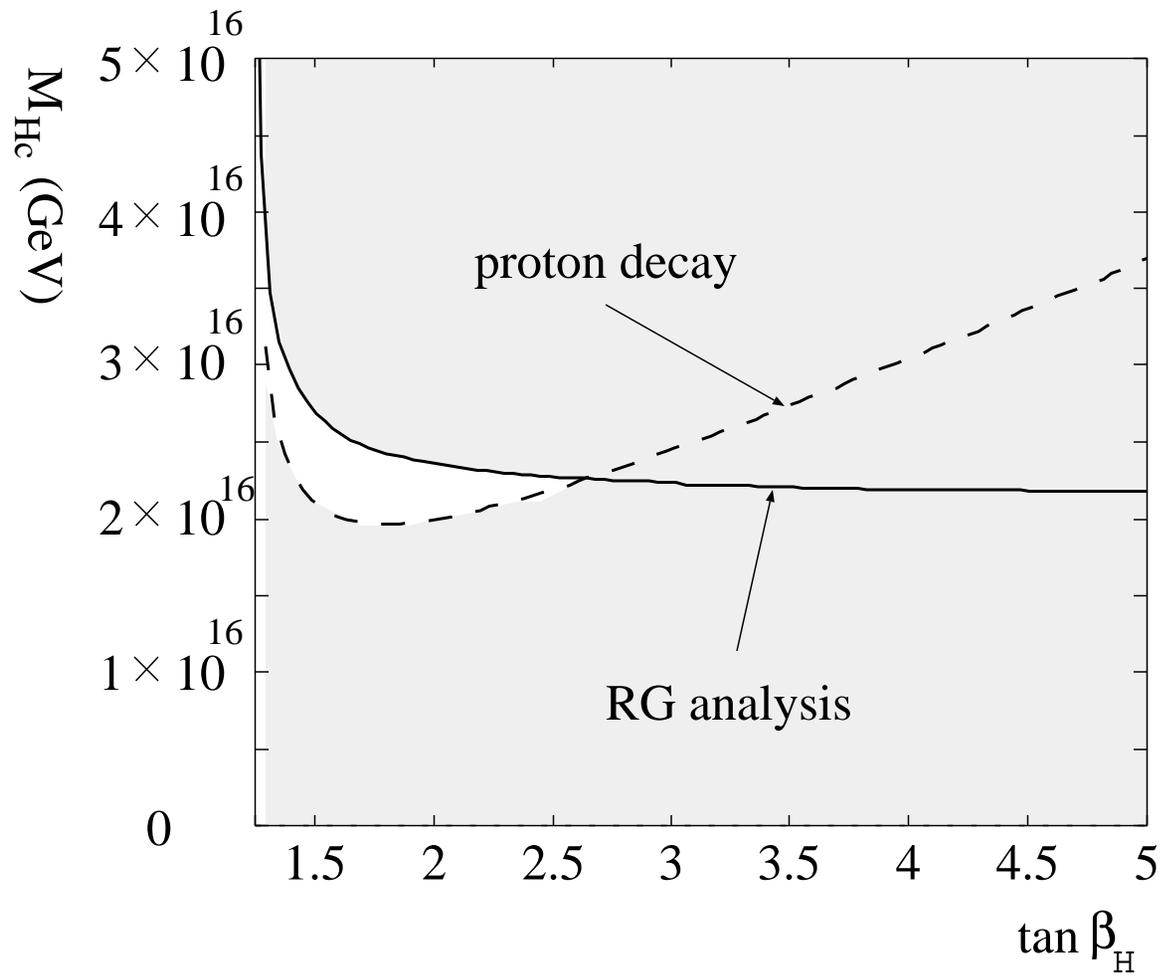
{\bf Fig.~2}\\ The upper limit on $\mhc$ from the RG analysis and the
lower limit from the negative search for the proton decay as a function
of $\tan\beta_{H}$. The solid line is the upper limit and the dashed
line the lower limit. The shaded region is excluded. Here, we take
$m_{\tilde q}=m_{\tilde l}=m_{\tilde h}=m_{H}=1~\TEV$,
$\beta=0.004~\GEV^{3}$, $m_{t}=174~\GEV$ and $\mwino=100~\GEV$.
\end{document}